\documentclass[preprint2]{aastex}

\slugcomment{Submitted to ApJ}

\newcommand{\ltsima} {$\; \buildrel < \over \sim \;$}
\newcommand{\gtsima} {$\; \buildrel > \over \sim \;$}
\newcommand{\lta} {\lower.5ex\hbox{\ltsima}}
\newcommand{\gta} {\lower.5ex\hbox{\gtsima}}
\def\flux{\rm\;erg\;s^{-1}\;cm^{-2}}

\shorttitle{NGC 6251}
\shortauthors{Chiaberge et al.}

\begin{document}

\title{The nuclear SED of NGC~6251: a  BL Lac in the center of an FR~I
radio  galaxy\thanks{Based  on  observations  obtained  at  the  Space
Telescope Science  Institute, which is operated by  the Association of
Universities  for  Research  in  Astronomy, Incorporated,  under  NASA
contract NAS 5-26555.}}


\author{Marco Chiaberge\altaffilmark{2}}
\affil{Space Telescope Science Institute, 3700 San Martin Drive,
Baltimore, MD 21218}
\affil{Istituto di Radioastronomia del  CNR - Via P. Gobetti
101, I-40129 Bologna, Italy}
\email{chiab@stsci.edu}

\author{Roberto Gilli}
\affil{Istituto Nazionale di Astrofisica (INAF) -- Osservatorio Astrofisico 
di Arcetri, Largo E. Fermi 5, I-50125 Firenze, Italy}
\author{Alessandro Capetti}
\affil{Istituto Nazionale di Astrofisica (INAF) -- Osservatorio  
Astronomico di  Torino,  Strada Osservatorio  20, I-10025   
Pino   Torinese,   Italy}   
\author{F.~Duccio Macchetto\altaffilmark{3}}
\affil{Space Telescope Science Institute, 3700 San Martin Drive,
Baltimore, MD 21218}


\altaffiltext{2}{ESA fellow}
\altaffiltext{3}{On assignment from ESA}


\begin{abstract}

We determine  the nuclear spectral energy distribution  (SED) from the
radio to  the gamma--ray band for  the FR~I radio  galaxy NGC~6251, by
using  both data  from  the  literature and  analyzing  HST and  X-ray
archival data.  In the $\log \nu - \log (\nu F_\nu)$ representation, the
SED has  two broad  peaks, and  it is remarkably  similar to  those of
blazars. We show that the low-energy peak can be explained in terms of
synchrotron radiation,  while the high  energy peak is  most plausibly
produced by inverse Compton scattering.  This brings direct support to
the FR~I--BL Lacs unification model.  We model the overall emission in
the frame of a synchrotron self-Compton scenario, which well describes
the  SED of  BL Lacs.   The model  parameters we  obtain  confirm also
quantitatively the  FR~I--BL Lac unification model and  imply a rather
small viewing angle to this source ($\theta \lta 20^\circ$).  NGC~6251
is the  second radio galaxy, in  addition to Centaurus~A,  for which a
similar analysis  has been performed.  A  significant improvement with
respect  to  the  case of  Cen~A  is  the  absence of  obscuration  in
NGC~6251, which strengthens the overall result.

\end{abstract}




\section{Introduction}

The basic  scheme for unification of  low-luminosity radio-loud active
galactic nuclei (AGN) assume that radio galaxies of the Fanaroff-Riley
class~I (FR~I, Fanaroff \& Riley  1974) and BL Lac objects differ only
because of a different viewing  angle to the relativistic jet (Urry \&
Padovani 1995  for a review).   In BL Lacs,  the jet is seen  at small
angles to the line-of-sight,  and the non-thermal emission produced at
its very  innermost base, on  the scale of $\sim  10^{15}-10^{16}$ cm,
dominates  the observed  radiation.  In  FR~I radio  galaxies  the jet
points at larger angles, and the ``BL Lac'' component should be beamed
away from the observer.

Clearly, the most direct implication of the unified models is that the
extended properties (e.g.  radio and optical emission, environment) of
the beamed sources and  their putative parent population should appear
the same.   Therefore, in the absence  of any direct  detection of the
nuclear  emission in  FR~I,  the  usual way  these  models are  tested
involves  the comparison  of  such extended  components  (e.g Urry  et
al. 2000, Antonucci \& Ulvestad 1985, Kollgaard et al. 1996).

However, faint unresolved optical  nuclei have been discovered in FR~I
radio galaxies  through HST observations  of a complete sample  of 3CR
sources \citep{pap1}.   The presence  of a strong  correlation between
these  optical  nuclei  and  the  radio  cores  argues  for  a  common
non-thermal  synchrotron origin of  both components.   Therefore, this
discovery  provides qualitative support  for the  low-luminosity radio
loud AGN unification models.

A more quantitative  analysis showed that cores of  radio galaxies are
overluminous with respect to what is expected from mis-oriented BL Lac
jets, for typical values of  the bulk Lorentz factor \citep{pap3}.  It
is therefore  plausible that the  optical cores of  radio galaxies
and the  radiation observed in  BL Lacs are  not produced in  the same
region of the relativistic jets: in BL Lacs, a ``fast jet spine'' with
Lorentz factors of  $\sim 15-20$ dominates, while in  FR~I we might be
observing emission from a slower (but still relativistic) shear layer.

Modeling  of  the overall nuclear  SED  could  provide an  even  more
stringent test for the unified  scheme, as it represents a fundamental
tool  for obtaining  information  on the  physical  conditions of  the
emitting region, as  it is usually done for  blazars (e.g.  Ghisellini
et al.   1998).  Unfortunately, the nuclear SED  of radio-galaxies are
in  general not  sufficiently  well sampled  to  follow directly  this
approach  and  radio  galaxies  are  generally  not  detected  in  the
gamma-ray band.

The first  radio galaxy (and, to  the best of our  knowledge, the only
case to  date) for  which a detailed  modeling of the  overall nuclear
SED,  from the radio  to the  gamma ray-band,  has been  undertaken is
Centaurus  A, the closest  radio galaxy  \citep{cena}.  The  result of
such  work is  that,  when considered  in  the comoving  frame of  the
source, the nuclear  SED of Cen~A can be  reproduced adopting the same
emission model and physical parameters  typical of BL Lacs of the same
total power, strongly supporting the unification scheme.

In this paper we focus on NGC~6251, which is a well studied FR~I radio
galaxy with a well collimated  one--sided radio jet  and a large scale
emission. Its total radio power is $L_{178} = 1.2 \times 10^{32}$
erg s$^{-1}$  (Waggett et al.   1977), i.e.   close to the  FR~I/FR~II
transition.   However,  its radio  morphology  is well
representative of the FR~I class.   The projected  linear size of the
extended  radio  structure   is 1.7  Mpc  (for  $H_0=75$   Km s$^{-1}$
Mpc$^{-1}$),  which  makes NGC~6251  one   of the largest known  radio
galaxies \cite{perley}.  The radio  source is associated with a  giant
elliptical galaxy, which hosts a $4-8 \times 10^{8} M_\sun$ black hole
\cite{ferrarese}.

By taking advantage  of the recent discovery that  the EGRET gamma-ray
source  J1621+8203  is  associated  with  NGC~6251  \citep{egret},  we
perform  a  detailed  analysis  of  its  overall SED  as  we  did  for
Centaurus~A. Our  result is that  the nucleus of this  low-power radio
galaxy can also be interpreted as a misoriented BL Lac.

The paper  is organized as follows:  in Sect.  2 we  describe the HST,
{\it Beppo}SAX and {\it Chandra}  data and the analysis procedures. In
Sect. 3 we determine the SED; in Sect. 4 we  discuss the nature of the
IR-to-UV  nuclear source,  which  is  crucial in  order  to model  the
overall SED.  In Sect. 5 we model the SED in the frame of a synchrotron
self-Compton scenario, and we  discuss the resulting model parameters,
while in Sect.  6 we draw conclusions.

A redshift of $z=0.0249$ for NGC~6251 (NED) is adopted throughout this
paper.

\section{Optical and X-ray data}

\begin{figure*}
\label{gallery}
\epsscale{2.2} 
\plotone{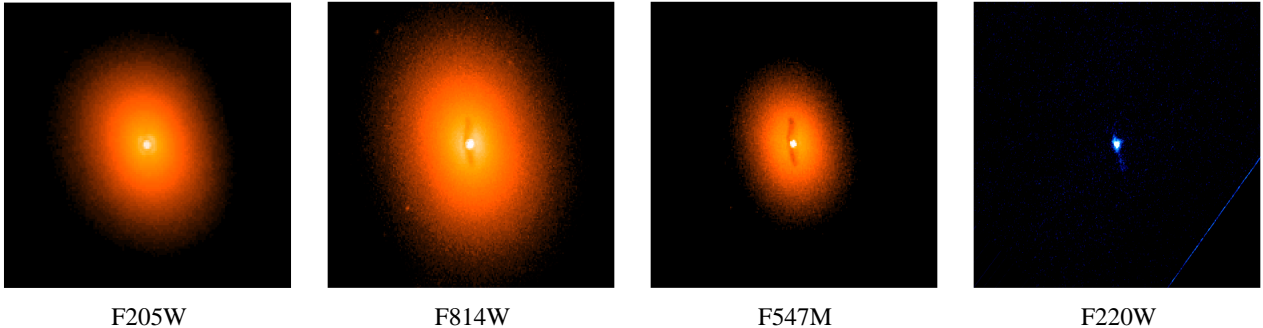}
\caption{The central regions of NGC~6251 as they appear in HST images,
from the IR to the UV band.  The pivot  wavelengths of the filters are
16060 \AA, 8336 \AA,  5461 \AA, 2329  \AA~~(Biretta et al. 2002).  The
projected  size    of    the   images    is   $10^{\prime\prime}\times
10^{\prime\prime}$ for all images except for  the UV (F220W) FOC image
which it is $6^{\prime\prime}\times 6^{\prime\prime}$.}
\end{figure*}

In order to  determine the SED, we first focus on  the analysis of the
IR-to-UV band and X-ray band observations.  These spectral regions are
crucial in order to model the observed radiation, since the data allow
us to set the position of the emission peaks, i.e.  the frequencies at
which most of the energy is released by the source, and thus constrain
the model.

\subsection{HST data}
\label{hstdata}

We have  measured the flux of  the central compact  source in archival
HST images, from the IR to the UV. In Table \ref{HSTobs} we report the
details  of the  observations, and  in  Fig. \ref{gallery}  we show  a
picture gallery of  the nuclear regions of NGC~6251  as they appear in
the images. The  F205W and F555W filters might  have contribution from
emission lines which  is however expected to be small  (a few \%), due
to their small equivalent width compared with the width of the filters
passband.   The  other  filters   are  centered  in  spectral  regions
relatively free of strong emission lines. Therefore (with the possible
exception  of  F205W and  F555W)  the  flux  we measure  is  continuum
emission.

The  procedure  is  as   follows.   We  perform  aperture  photometry,
evaluating the total counts in a circular region centered on the point
source,  with  different radii  for  the  different  filters. This  is
important in order to have  a reliable determination of the total flux
for the  point source  allowing for slightly  different shapes  of the
point spread function.  For the  FOC UV data both the determination of
the   background  and   the   choice  of   the   aperture  radius   is
straightforward, since the nucleus is by far the dominant component in
the  image. For  the WFPC2  images, due  to the  presence of  the host
galaxy stellar  component and an  extended dusty disk  which surrounds
the  unresolved   nucleus,  the  largest   source  of  error   is  the
determination of the background. Thus  we have adopted the rather well
tested  procedure  of measuring  the  flux in  a  radius  of 5  pixels
(corresponding  to  $0.2$ arcsec),  measuring  the  background in  the
surrounding  annulus (1 pixel  width).  This  is essentially  the same
procedure as we  used for the analysis of the  complete samples of 3CR
radio galaxies \citep{pap1,pap4}.

Due to  the different shape of  the PSF in the  different filters, the
lower resolution  at IR wavelengths and  the larger pixel  size of the
detector, the NICMOS  data have to be corrected for  PSF loss. We have
produced synthetic  PSFs with  the TINYTIM software  \citep{krist} for
the  different instrument  configurations.  We  compared  the observed
shape of the nuclear PSF in NGC~6251 with the respective synthetic PSF
and  we selected an  appropriate aperture  radius for  the photometry
($r=5.5$ pixels, $r=4$  pixels, $r=3$ pixels for the  F205W, F160W and
F110W filters, respectively).  This  procedure ensures that we measure
the background as  close as possible to the  nucleus without suffering
contamination from  the nucleus  itself.  However, we  have determined
that  in this  case the  aperture  photometry radius  includes only  a
fraction  of the  total counts.  Therefore, in  order to  estimate the
total flux of the point source, we have derived the renormalization
factors from the analysis  of the synthetic PSFs ($k=1.234$, $k=1.37$,
$k=1.38$ for the F205W, F160W  and F110W filters, respectively) and we
have corrected the measured flux accordingly.

The  effective frequencies  of the  observations have  been calculated
using SYNPHOT, by taking into  consideration the observed slope of the
continuum emission. The count rates  have been multiplied by the value
of  the  header  parameter  PHOTFLAM,  which  represents  the  inverse
sensitivity on the filters in the case of a flat continuum ($F_\lambda
\propto \lambda^\beta$, $\beta=0$, which corresponds to $F_\nu \propto
\nu^{-\alpha}$, $\alpha=2$).   We have tested that  the observed slope
does  not affect  the  values of  PHOTFLAM   significantly (less  than
$1\%$).  The fluxes measured in  the HST observations are reported  in
Table \ref{hstphot}. We estimate that the typical error on the nuclear
flux measurements is \lta $10\%$. The IR-to-UV slope is $\alpha=1.75\pm
0.16$, calculated between 16000 \AA~ and 2200 \AA.

\begin{deluxetable}{l l l c c}
\tablewidth{0pt}  
\tablecaption{HST observations}  
\tablehead{ \colhead{Instrument} &   \colhead{camera/config} &   
\colhead{Filter} &   \colhead{HST Prop. ID} &   \colhead{Obs. Date}}
\startdata
NICMOS & NIC2 & F205W  & GO7862 &  1998 Jul 06 \\
NICMOS & NIC2 & F160W  & GO7862 &  1998 Jul 06 \\
NICMOS & NIC2 & F110W  & GO7862 &  1998 Jul 06 \\
WFPC2  & PC   & F814W  & GO6276 &  1995 Jun 28 \\
WFPC2  & PC   & F555W  & GO6276 &  1995 Jun 28 \\
WFPC2  & PC   & F547M  & GO6653 &  1996 Sep 13 \\
FOC    & f/96 & F342W  & GO6246 &  1996 Feb 19 \\
FOC    & f/96 & F220W  & GO6891 &  1997 Jul 07 \\
\enddata
\label{HSTobs}
\end{deluxetable}

\begin{deluxetable}{l c c}
\tablewidth{0pt}  
\tablecaption{HST Photometry}  
\tablehead{\colhead{Filter} &\colhead{Eff. Wav.} &   \colhead{Flux}\\
\colhead{~} & \colhead{\AA} &   \colhead{erg cm$^{-2}$ s$^{-1}$ \AA$^{-1}$ }}
\startdata
F205W & 21063 &   $7.3\times 10^{-17}$ \\
F160W & 16250 &   $6.2\times 10^{-17}$ \\
F110W & 11777 &   $6.8\times 10^{-17}$ \\
F814W & 8124  &   $9.1\times 10^{-17}$ \\
F555W & 5558  &   $1.3\times 10^{-16}$ \\
F547M & 5501  &   $1.2\times 10^{-16}$ \\
F342W & 3435  &   $8.4\times 10^{-17}$ \\
F220W & 2414  &   $9.1\times 10^{-17}$ \\
\enddata
\label{hstphot}
\end{deluxetable}

\subsection{X-ray observations and results}
\label{xrays}

We have  analyzed unpublished recent {\it Beppo}SAX  and {\it Chandra}
X-ray data.  These observations are available in the respective public
archives. We  focus in particular on the  {\it Beppo}SAX observations,
since NGC~6251 is known to be  a bright X-ray source (2-10 keV flux of
$\sim 10^{-12}\flux$;  Sambruna, et al.  1999) and therefore  the {\it
Chandra} data  are significantly affected  by pileup.  However,  it is
extremely useful  to compare the  two observations.  Given  its arcsec
spatial resolution,  which is  $\sim 200$ times  sharper than  that of
BeppoSAX, {\it Chandra} is indeed the ideal experiment to pinpoint the
nuclear  emission.   In Sect.   \ref{chandra}  we  show  that the  two
observations have consistent results,  once that the pileup effects in
{\it Chandra} data are taken into account.

\subsubsection{BeppoSAX data}

NGC~6251  was  observed  in  July  2001  by the  LECS,  MECS  and  PDS
instruments on board  BeppoSAX (with exposure times of  30.2, 80.9 and
36.2  ksec, respectively),  as part  of  an AO5  GO observation  (code
51322001, P.I. M. Guainazzi).  The LECS and MECS proportional counters
are  operating in  the 0.1-10  keV and  1-10 keV  bands, respectively,
while the PDS instrument is  a scintillation detector operating in the
high energy  band 15-300  keV.  In our  spectral analysis  we consider
only  the  range  of  energies  in  which  the  instruments  are  best
calibrated (i.e.  0.12-4 keV for the LECS, 1.6-10 keV for the MECS and
15-220 keV for the PDS).

We perform data analysis by  using standard procedures. We extract the
source spectrum from an aperture of  8 and 4 arcmin radius in the LECS
and  MECS images, respectively,  and we  evaluate the  background from
blank  sky fields.   The PDS  data reduction  is as  performed  by the
archive pipeline.  We  perform the spectral analysis of  MECS and LECS
data by using XSPEC v11.0 on  rebinned spectra with at least 20 counts
per bin to use the $\chi^2$-statistics in the fitting process.  Errors
are given at the 90\% confidence level for one interesting parameter.
The  cross-calibration constant  between  LECS and  MECS, obtained  by
fitting  simultaneously the  LECS and  MECS  data in  the common  band
1.6-4.0  keV, is  found to  be 0.67.   The  cross-calibration constant
between  the MECS  and  the PDS,  constrained  to be  in the  fiducial
range\footnote{see the  BeppoSAX handbook (Fiore,  Guainazzi \& Grandi,
1999,   Cookbook   for    BeppoSAX   NFI   Spectral   Analysis)   {\tt
http://www.asdc.asi.it/bepposax/software/index.html}}   0.77-0.93,  is
found to be 0.88.  Given the  PDS large field of view (78 arcmin HWZI)
and lack of  spatial resolution, other sources of  high energy photons
might  in principle  contaminate the  PDS data.   We search  for X-ray
sources  in the  MECS  image (25  arcmin  radius FOV)  and, at  larger
offsets,     in     the     ROSAT     source     catalogs\footnote{\tt
http://www.xray.mpe.mpg.de/cgi-bin/rosat/src-browser}.   Based on the
fluxes  and  spectra of  the  X-ray sources  within  the  PDS FOV,  we
estimate  that more than  $90\%$ of  the PDS  photons are  produced by
NGC~6251.

The 0.1--200 keV spectrum  is well  described ($\chi^2/dof=180.5/183$)
by a  power  law  with  a   photon index  $\Gamma=1.79\pm0.06$  mildly
absorbed by  a  column density   $N_H=4.7^{+4.3}_{-4.7}\times 10^{20}$
cm$^{-2}$   in excess   of  the  Galactic   value ($N_{HGal}=5.5\times
10^{20}$ cm$^{-2}$; Dickey \& Lockman 1990). Similar results have also
been found by Guainazzi et  al. (2003).  Since  the nuclear source  is
embedded in a diffuse gas halo  (e.g.  Birkinshaw \& Worrall 1993), we
add in the   spectral fit a  thermal component  using the {\tt  mekal}
model  of XSPEC.   In  this  case, we  obtain  a comparably  good  fit
($\chi^2/dof=172.6/180$),    although  the estimate    of the  nuclear
absorption   rises to $N_H=1.4^{+1.2}_{-0.7}\times 10^{22}$ cm$^{-2}$,
while   the   power-law      spectral    index   remains     unchanged
($\Gamma=1.78^{+0.13}_{-0.34}$).    The thermal component  is found to
have  a  temperature   $kT=1.20^{+0.97}_{-0.34}$ keV,   with  a  metal
abundance $Y=0.09^{+0.22}_{-0.06}$.    The  latter estimate    of  the
nuclear absorption, the power--law spectral index, as well as the halo
temperature and metal  abundance, are in  good agreement with the 1994
ASCA     observation    \citep{sambruna99}\footnote{\citet{sambruna99}
obtained:      $N_H=7.5^{+6.4}_{-5.8} \times     10^{21}$   cm$^{-2}$,
$\Gamma=1.83^{+0.21}_{-0.18}$,       $kT=1.04^{+0.21}_{-0.18}$    keV,
$Y=0.06^{+0.51}_{-0.04}$, while a somewhat  steeper powerlaw and lower
absorption have been found  by Turner et al.  (1997) in the same  ASCA
dataset}.  However, we point out that the 2-10 keV {\it Beppo}SAX flux
($4.78\pm0.07 \times 10^{-12}\flux$)  is significantly higher than the
flux  measured     in  the   ASCA   observation  ($1.40\pm0.04  \times
10^{-12}\flux$), suggesting that  the source flux  is variable in  the
X-rays.

\subsubsection{Chandra data}
\label{chandra}

NGC~6251 was observed by Chandra (37.4 ksec  with ACIS-I) in September
2000, as  part of an AO1  GO project (Obs.  ID  847, P.I.   Kerp). The
source is 3 arcmin away  from the telescope  aimpoint and falls in the
gap among the  4 ACIS-I CCDs, where  the  exposure map  is  low.  This
reduces  the  number of detected  photons,  but  does not mitigate the
pileup effects. Based on the BeppoSAX spectrum  and flux, we estimated
that in  the Chandra observation (with  a  frametime of  3.24 sec) the
pileup fraction is above 60\% (the pileup  fraction decreases to $\sim
20\%$ if the ASCA  flux  is used).  We used   CIAO v2.2 to perform   a
standard data reduction and XSPEC v11.0 for the spectral analysis.  As
expected, due  to  the high  pileup fraction, the  ACIS-I  spectrum of
NGC~6251 is   much harder  than  that observed  by BeppoSAX,  having a
photon index    $\Gamma=1.30\pm0.26$.     By  using  the    MARX   4.0
simulator\footnote{Wise M.W.   Davis, J.E., Huenemoerder  D.P.,  Houck
J.C.,  Dewey,   D.,    2003,   MARX   4.0    Technical  Manual:   {\tt
http://space.mit.edu/CXC/MARX}}  we verified  that  both the  spectral
slope and  the  flux are  in excellent   agreement   with the  Chandra
observation,  under the  assumption  that the  input spectrum  is that
observed by BeppoSAX.   We point out that it  is appropriate to compare
such observations, since the {\it Beppo}SAX spectrum was obtained much
closer in time to the Chandra data than the ASCA observation.

\section{The nuclear Spectral Energy Distribution of NGC 6251}

\begin{figure}[h]
\epsscale{1} \plotone{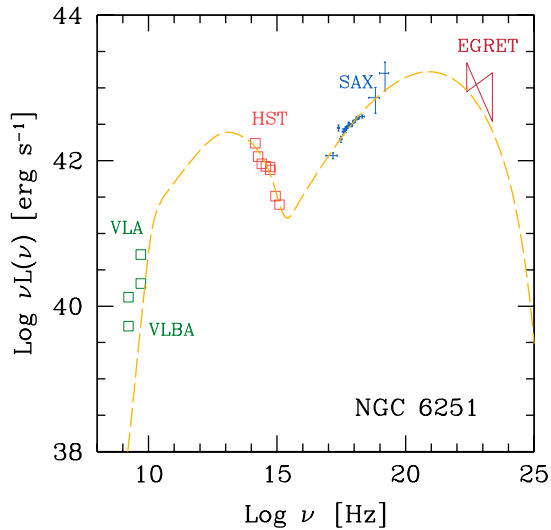}
\caption{The  overall spectral energy  distribution of  NGC~6251, from
the  radio to  the gamma-ray  band.  Details of  the data  are in  the
text.  The dashed  line  corresponds to  the  SSC model  we derive  in
Sect. \ref{modelling}.}
\label{6251sed}
\end{figure}

In  Fig.~\ref{6251sed} we show  the overall  nuclear SED  of NGC~6251.
The radio data are taken  from \citet{jones86}. The VLBA fluxes of the
unresolved radio core  are a factor of 2.5 lower than  the VLA both at
6cm  and 18cm.   The IR-through-UV  HST  data and  the {\it  Beppo}SAX
X--ray  observations   and  data  analysis  have   been  discussed  in
Sect.\ref{hstdata} and  Section \ref{xrays}, respectively.   The EGRET
spectrum  is the  fit to  the spectrum  of the  source, as  taken from
\citet{egret}.  The SED in a $\log \nu- \log \nu F_\nu$ representation
has two  broad peaks and, although  the peak frequencies  are not well
constrained,  they are  clearly located  in the  far IR-mm  region and
between the X-rays and gamma-rays.

As in the case of Centaurus~A \citep{cena}, the SED appears remarkably
similar to the typical ``blazar--like''  SED. In such objects, the low
and  high energy  peaks are  believed  to be  produced by  non-thermal
synchrotron  and  inverse-Compton  radiation,  respectively.   In  the
following we discuss in detail  the origin of the optical emission and
we show that  the SSC emission model can also  account for the nuclear
radiation of NGC~6251.

\section{The nature of the active nucleus in NGC~6251 and the 
origin of the optical nuclear emission}

\begin{figure}[h]
\epsscale{1} \plotone{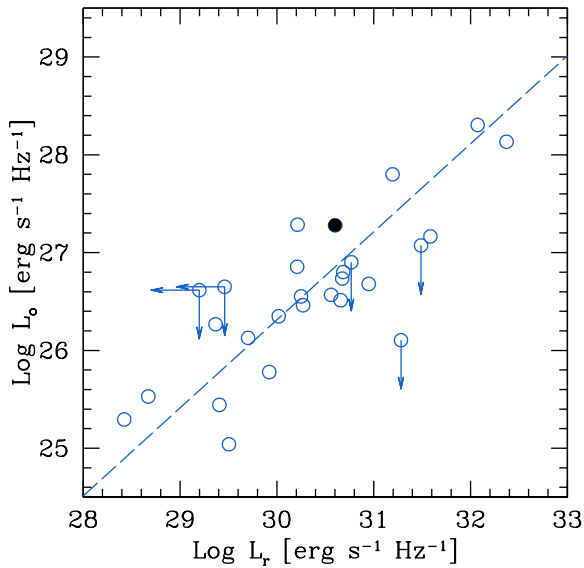}
\caption{The  correlation  between the  radio  core  at  5GHz and  the
optical (R band) cores seen in HST images, for the FR~I radio galaxies
of  the   3CR  sample  \citep{pap1}.   The  black   filled  circle  is
representative of NGC~6251.}
\label{lum6251}
\end{figure}

There is growing  evidence that the nuclei of  FR~I radio galaxies are
basically unobscured, and the  compact nuclei observed in the majority
of  FR~I  are  non-thermal  radiation  produced at  the  base  of  the
relativistic jet.   The main  evidence for this  is the presence  of a
strong linear correlation between FR~I's optical and the radio nuclear
radiation,   the  latter  being   definitely  of   non-thermal  origin
\citep{pap1,verdoes,b2}.   Although the general  behavior of  the FR~I
class  as a whole  seems to  be understood,  for ``single  sources'' a
certain ambiguity on  the nature of the optical  nuclei might still be
present.   Since   the  IR  through   UV  radiation  is   crucial  for
constraining  the  frequency  of  the  synchrotron peak,  it  is  very
important  to assess  its nature.   On the  other hand,  for  the hard
X--ray and  gamma ray  emission the interpretation  as inverse-Compton
emission appears to be the only viable possibility.

The origin of  the bright optical core of  NGC~6251 has been discussed
by  \citet{ferrarese}.   There are  two  scenarios  proposed by  these
authors: a non--thermal process (most plausibly synchrotron emission),
or thermal radiation from an optically thick accretion disk.

Let us  consider the position  of NGC~6251 in diagnostic  diagrams for
the  nuclei  of  radio  galaxies.   In Fig.~\ref{lum6251}  we  plot
NGC~6251 in the plane formed  by the luminosity of the optical nucleus
of versus  the radio core.   The open circles  are the nuclei  of FR~I
radio galaxies from the 3CR  catalog, as taken from \citet{pap1}.  The
dashed line  is the regression  line between the two  quantities.  The
black  point,   representative  of  NGC~6251,  lies   just  above  the
correlation.  This is  a strong clue that NGC~6251  behaves exactly as
other  FR~I.  We stress  that the  comparison with  the 3CR  sample is
extremely appropriate, since  \citet{laing}, have included NGC~6251 in
their complete revised 3CR sample (the 3CRR).

\begin{figure*}
\epsscale{1.8} 
\plottwo{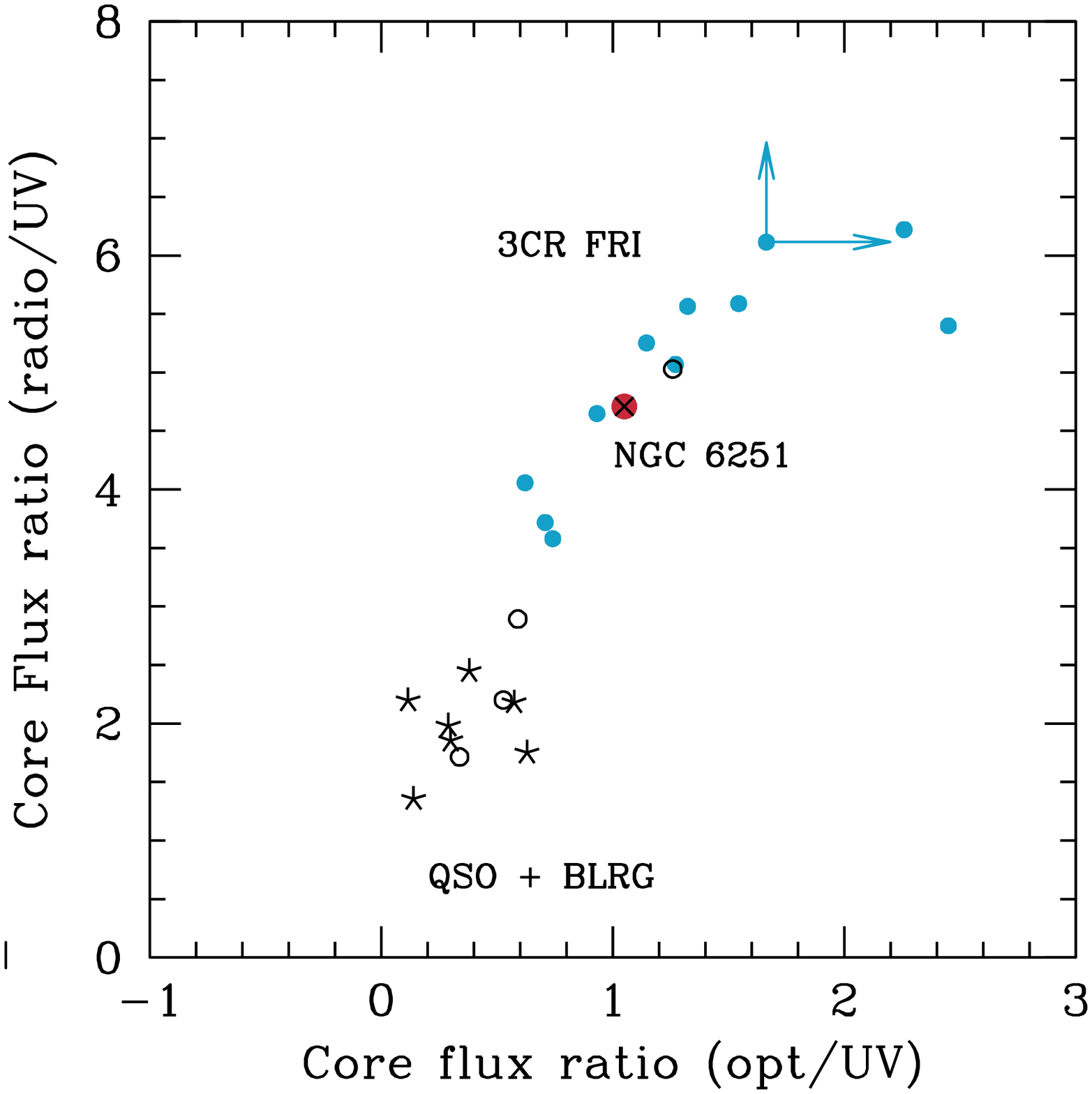}{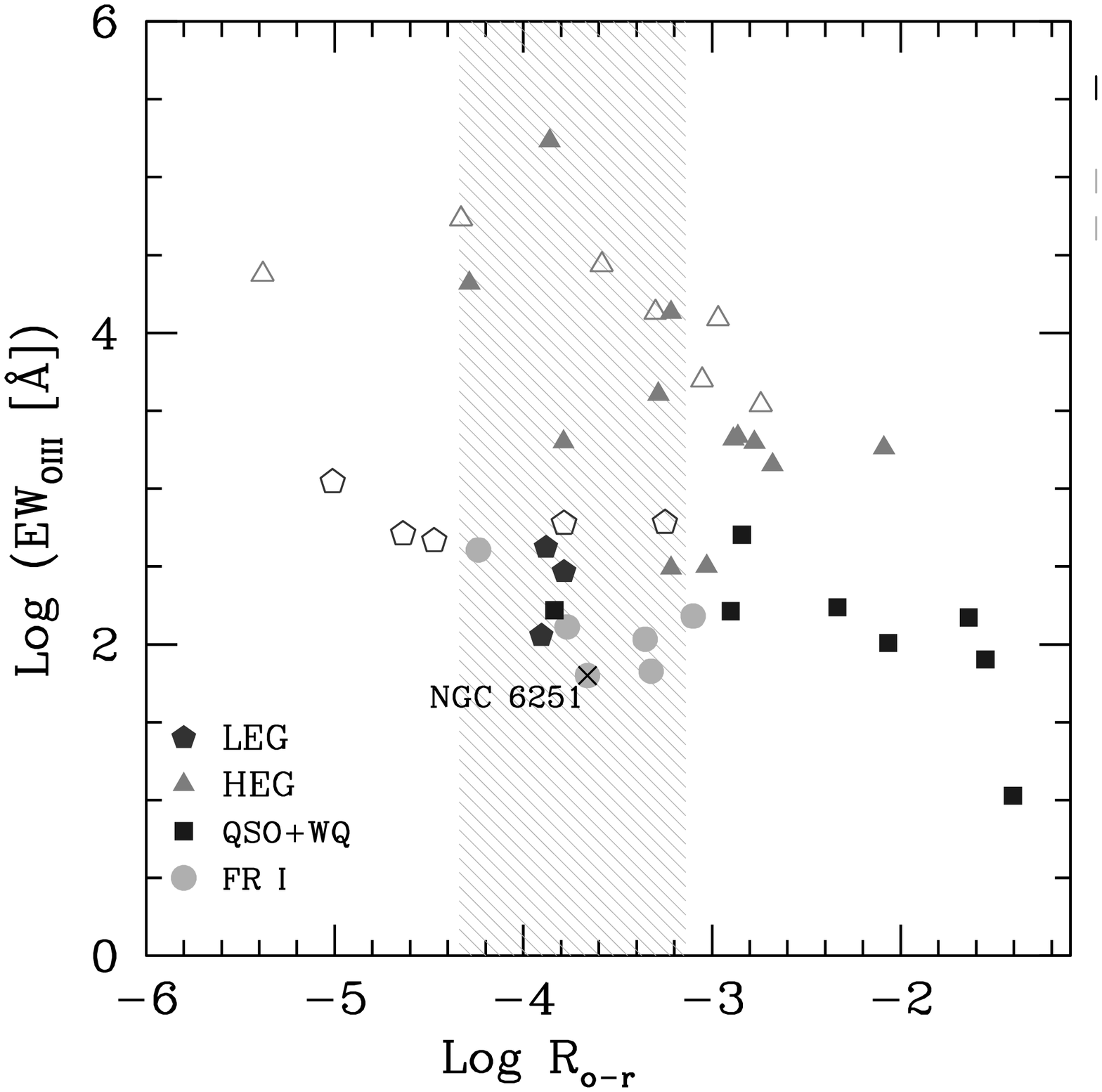}
\caption{{\it  Left  panel  (a):}  Radio core/UV  nuclear  flux  ratio
vs. optical/UV nuclear flux ratio for 3CR FR~I (filled circles), FR~II
(empty circles) and radio loud quasars with $z<0.3$ from the sample of
\citet{elvis}.  The  empty circle  that lies among  FR~I is  3C~338, a
low-excitation   galaxy   (see   discussion   in  Chiaberge   et   al.
2002b).  {\it Right  panel (b):}  Equivalent width  of  the [OIII]5007
emission line (measured with  respect to the optical nuclear emission)
plotted vs.  the ratio between  the optical nuclei to radio core flux,
for  a sample  of 3CR  FR~I  and FR~II  \citep{pap4}.  Filled  symbols
represent detections,  empty symbols are  upper limits to  the optical
nuclear  emission.   Different  symbols  correspond to  the  different
subclasses  of  radio  galaxies.    The  shaded  area  represents  the
dispersion (1$\sigma$)  of the  linear correlation between  radio core
and optical nuclear luminosity found for FR~I.}
\label{diagnostics}
\end{figure*}

In Fig.   \ref{diagnostics}a we  show the position  of the  nucleus of
NGC~6251 in the plane formed  by the radio-to-UV vs. the optical-to-UV
flux  ratios.  As  discussed  in \citet{papuv},  the  position of  the
objects  in this  plane is  sensitive both  to absorption  and  to the
different intrinsic spectral properties of the nuclei.  Steep spectrum
radio quasars (plotted as stars) and broad lined radio galaxies (empty
circles)   occupy  the  same   region  of   the  plane,   and  clearly
differentiate from FR~I mainly because  of the presence of the thermal
blue bump in the optical-to-UV  spectral range. NGC~6251 again lies in
the region of FR~I.

However,  as already  pointed  out above,  for  ``single objects''  we
cannot exclude on  the basis of the radio-optical  properties that the
optical  emission could be  scattered light  produced by  an accretion
disk  obscured to  our  line-of-sight.   In this  case,  we know  from
\cite{pap4}   that  its   representative  point   would  lie   on  the
radio-optical correlation of Fig.~\ref{lum6251} {\it by chance}.  This
has been indeed found to be  rather common among FR~II in which strong
narrow emission lines are observed.

In order to remove the degeneracy in the interpretation of the optical
emission, we use the [OIII]5007  emission line, which is indeed a good
indicator  of the  intrinsic  ionizing radiation  field.  From  ground
based   observations  ($2.7^{\prime\prime}   \times  4^{\prime\prime}$
aperture,  centered  on the  galaxy  nucleus)  the  luminosity of  the
[OIII]5007 line is $5  \times 10^{39}$ erg s$^{-1}$ \citep{shuder}.  A
useful diagnostic is  the ratio between the flux  of the emission line
and  the observed  nuclear continuum  emission (i.e.   the ``nuclear''
equivalent width).  This has been used to discriminate between objects
in which the nuclear source is  seen directly and objects in which the
ionizing radiation field is obscured to our line-of-sight and observed
only through  scattered radiation  \citep{pap4}.  In the  latter case,
high values  of the  equivalent width are  expected, since only  a few
percent  of  the  intrinsic  nuclear  source  is  observed.   In  Fig.
\ref{diagnostics}b we show  the position of the source in
the plane  formed by the  [OIII] EW plotted  against the ratio  of the
optical nuclear  flux to  radio core flux  $R_{\rm o-r}$.   The shaded
area  corresponds to the  dispersion ($1  \sigma$) of  the correlation
between these two quantities  for FR~I galaxies.  The equivalent width
of the [OIII]5007  emission line for NGC~6251 is  typical of the other
unobscured FR~I nuclei.  As  already known from Fig~\ref{lum6251}, the
value of $R_{\rm o-r}$ is also in the same region as the other FR~I.

In  the light of  the above  analysis, we  conclude that  the dominant
physical origin  for the  optical nucleus is  non--thermal synchrotron
emission  produced  at  the  base  of  the  jet.   Any  other  thermal
contribution to the observed flux should then be considered lower than
(or equal to) the observed synchrotron radiation.

\begin{figure}
\epsscale{1} 
\plotone{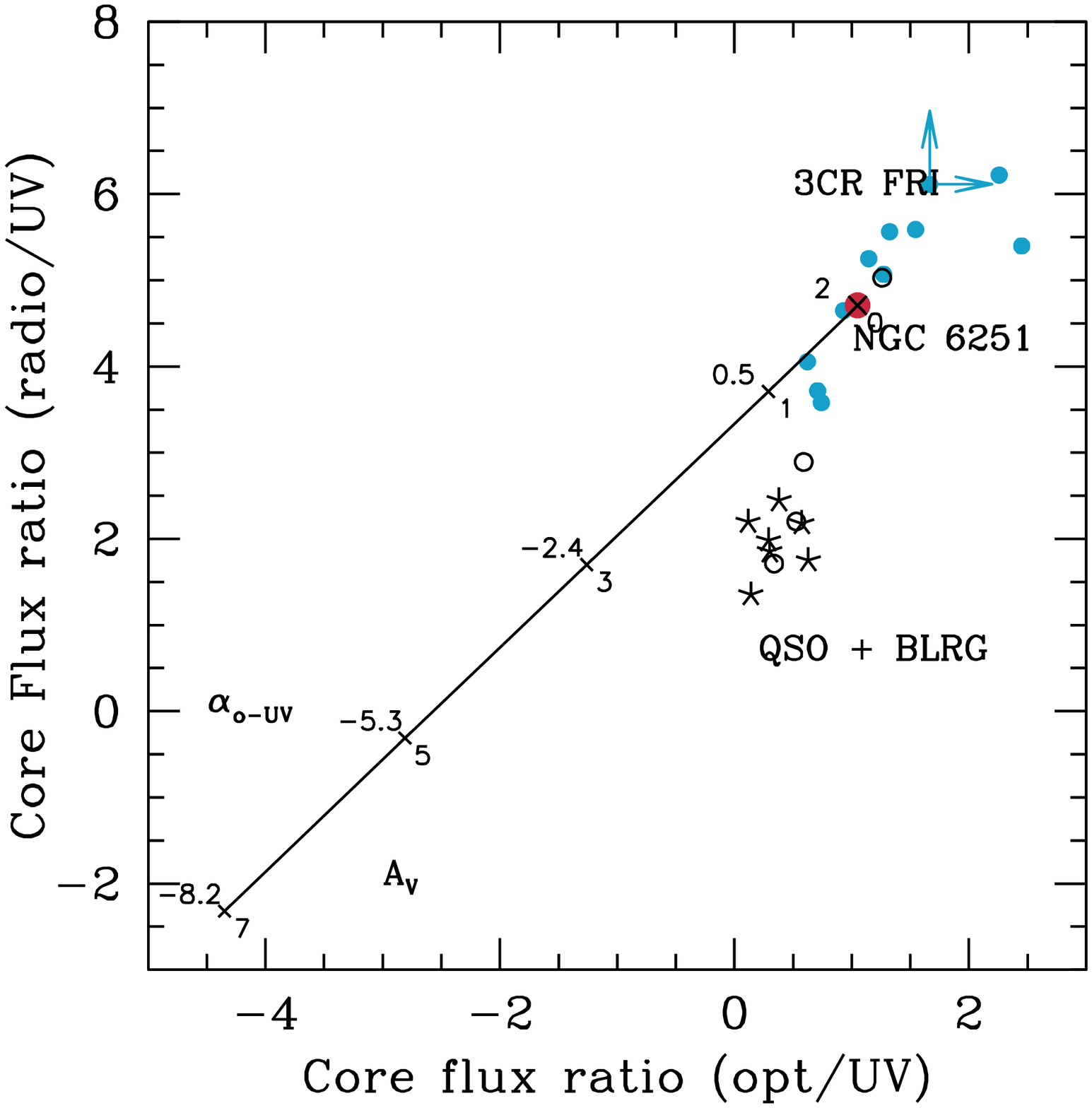}
\caption{Radio core/UV nuclear flux  ratio vs. optical/UV nuclear flux
ratio for 3CR  FR~I (filled circles), FR~II (empty  circles) and radio
loud quasars  with $z<0.3$,  as in Fig.~\ref{diagnostics}b.  The solid
line is the dereddening curve for NGC~6251 (see text for details).}
\label{uvtrace}
\end{figure}

\begin{figure}
\plotone{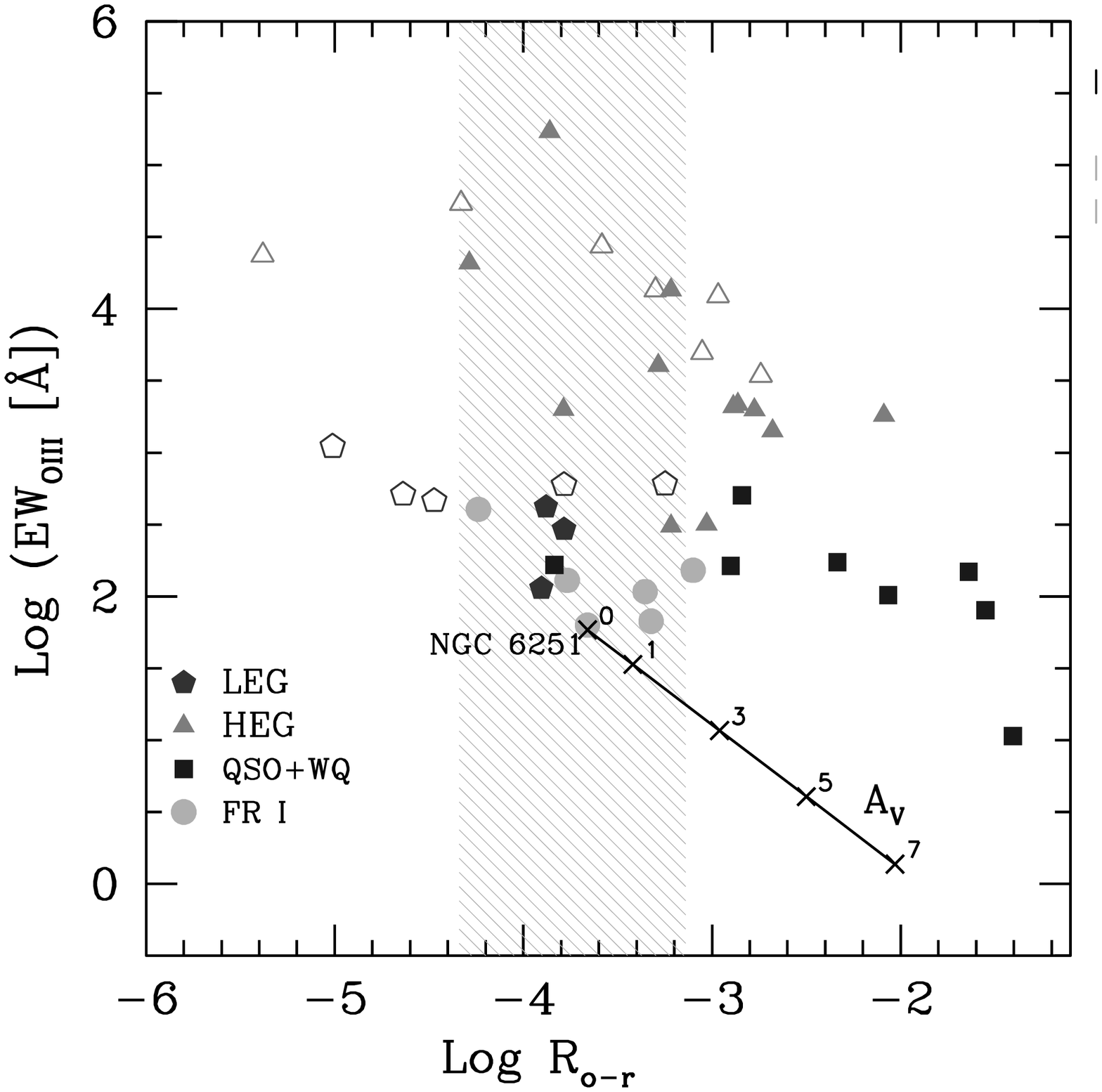}
\caption{Equivalent width of the  [OIII]5007 emission line plotted vs.
the  ratio between  the optical  nuclei to  radio core  flux,  for the
sample    of    3CR   FR~I    and    FR~II    \citep{pap4},   as    in
Fig. \ref{diagnostics}b.  The solid  line is the dereddening curve for
NGC~6251 (see text for details).}
\label{lo3}
\end{figure}

\subsection{What is the role of absorption in NGC~6251?}

Before modeling  the nuclear emission,  we must briefly  consider the
possible presence  of absorption on the line-of-sight  to the nucleus,
since one of the two models  for the X-ray emission includes a certain
amount of  $N_H$ in  excess of the  Galactic value.   Absorption would
indeed naturally steepen the IR-to-UV slope of the continuum emission,
and could in principle  ``artificially'' produce a two-peak SED.  This
is an  issue for objects  like Cen~A, because  of the presence  of the
extended  dust  lane which  certainly  absorbs  the nuclear  radiation
\citep{marconi}.  However, in NGC~6251 there is no apparent reason for
absorption in the optical.  Firstly, the dusty disk does not appear to
hide  the nucleus.   Although the  extinction  from the  disk is  only
$A_V\sim 0.6$mag  (in the external regions),  this would significantly
affect the  UV emission ($A_{2400  \rm \AA}\sim 2.5 A_V$,  Cardelli et
al.  1989). If  the disk were to cover the nucleus  we would expect to
see  a  progressive steepening  of  the  spectral  slope as  frequency
increases,   which  is   not  observed.    Furthermore,   the  typical
optical-to-UV  spectral  slope  of  FR~I  from the  3CR  catalog  lies
somewhere between $\alpha_{o-UV}=1-2$  \citep{papuv}, while only those
objects in which absorption features are apparent have steeper slopes.
The value of $\alpha_{o,UV} =  2.0\pm 0.2$ that we derive for NGC~6251
appears to be consistent with other {\it unobscured} FR~Is.

In Fig. \ref{uvtrace} we show the effects of a small amount of optical
extinction on  the broad--band spectral   properties of NGC~6251  when
compared with    other radio  galaxies.     We  have overplotted   the
de-reddening      line     for   NGC~6251      on      the plane    of
Fig.~\ref{diagnostics}a, assuming  a range of  $A_V$ from 0 to 7. This
confirms that only an $A_V  \lta 1-1.5$ mag   is allowed.  For  higher
values  of extinction, the broad  band spectral properties of NGC~6251
would be incompatible with any of the other sources  in the same range
of redshift and total power.  In particular, it appears that it is not
possible to ``drag'' the source into the region occupied by quasars if
we only consider the effects of obscuration.  Note that
since   the typical error  for each  single  flux measurement is $\lta
10\%$, the resulting error on the data plotted here is of the order of
the  size  of  the symbols. Therefore,  we  exclude  that the  nuclear
emission is thermal radiation from an accretion disk seen directly.

Similarly to  Fig. \ref{uvtrace} we have  overplotted the de-reddening
curve  on the  EW([OIII]) vs.   $R_{\rm o-r}$  diagnostic  plane (Fig.
\ref{lo3}).   An  $A_V  \gta  1-2$  mag would  again  move  the  point
representative of NGC~6251  in a region of the  plane where no sources
are found.

The fact that the X-ray data  can be modeled with absorption in excess
of the galactic  value, while in the optical there  is no evidence for
significant  extinction  is not  surprising.   In  fact, the  standard
conversion between  $N_H$ and $A_V$,  adopting local Galactic  dust to
gas  ratio  (which  in  our  case would  imply  $A_V=5\times  10^{-22}
N_H\simeq 7$ mag) generally  does not hold in AGNs.  \citet{maiolino1}
have found that in AGNs $E(B-V)/N_H$ is always lower than the Galactic
value by  a factor  of 3 to  100.  Several different  explanations for
such  a  discrepancy  have  been  proposed by  various  authors,  e.g.
non--standard values  for the  dust-to-gas ratio in  the ``absorbing''
structures,  a different location  for the  optical and  X-ray sources
and/or for  the absorbing  structures, or the  presence of  large dust
grains  which would  flatten  the extinction  curve and  significantly
lower the $A_V/N_H$ ratio \citep{maiolino2}.

As  noted  above,  from the  analysis  of  the  X-ray data  we  cannot
discriminate between the  two models that fit the  data.  However, for
our purpose,  the most  important parameter we  derive from  the X-ray
data  is the slope  of the  (non-thermal) power--law  component, which
remains unchanged in either models.

In the following we discuss the nature of the optical emission through
various diagnostics.  Then we proceed  to model the overall SED in the
synchrotron  self-Compton scenario,  with the  same method  as usually
done for blazars.

\section{SSC modeling of the SED}
\label{modelling}

\begin{deluxetable}{l  l}
\tablewidth{0pt}   \tablecaption{}
\tablehead{~~~~~~~~~~~~~~~~~~~~~~~~~~Model   parameters} 
\startdata
$R  = 5.5\times 10^{16}$ cm         &  $B =  0.03$ G \\
$L_{inj}= 4.0\times 10^{44}$ erg s$^{-1}$ &  $\delta =3.2$  \\
$\gamma_{min}= 4.0 \times 10^{3}$ &  $\gamma_{max}= 4.5 \times 10^{4}$ \\ 
$p= 1.9 $                         &  $t_{esc} = 15 R/c$  \\
\enddata
\label{modpar}
\end{deluxetable}

We model the SED of NGC~6251 in the frame of a homogeneous synchrotron
self-Compton scenario (SSC), which is successfully used to account for
the  overall emission  of  blazars (e.g.   Ghisellini  et al.   1998).
Details of  the model are given  in \citet{cg} and a  short summary is
reported  for  clarity  in  Appendix \ref{appendice}.   The  model  is
particularly appropriate for low-power objects (BL Lacs), in which the
external radiation field  is probably low, and the  dominant source of
seed  photons  for   inverse-Compton  scattering  is  the  synchrotron
radiation  field itself.   Since the  total radio  power  of NGC~6251,
($L_{178} = 1.2 \times 10^{32}$  erg s$^{-1}$, Waggett et al. 1977) is
in the  range spanned by BL Lacs,  and close to its  median value (see
e.g.   Perlman  et al.   1996,  Kollgaard  et  al.  1996,  Cassaro  et
al. 1999), the use of the SSC model appears to be appropriate.

In Table \ref{modpar}  we show the parameters of  the model which best
reproduce the SED. All of them  are typical of BL Lacs, except for the
value of  $\delta$, the relativistic  beaming factor\footnote{$\delta$
is defined as $\delta=1/\Gamma(1-\beta \cos\theta)$, where $\Gamma$ is
the  bulk  Lorentz factor  of  the jet  and  $\theta$  is the  viewing
angle.}.  In order to perform a more detailed analysis, we compare our
model parameters with the large BL Lac sample modeled by \citet{gg98}.
We use for comparison  the compactness parameter $\ell_{inj} = L_{inj}
\sigma_T /(R  m_e c^3)$,  which represents the  ratio of  the injected
power  to   the  size   of  the  source.    For  NGC~6251   we  obtain
$\ell_{inj}=0.2$, which lies at the  higher end of the distribution of
$\ell_{inj}$ for BL Lacs. Both the  magnetic field $B$ and the size of
the source are well inside  the range spanned by BL Lacs ($10^{-2}\lta
B_{BL Lac} \lta 1.2$G and $10^{16} \lta R_{BL Lac} \lta 10^{17}$cm).

As already  noted above, $\delta$ is  lower than in BL  Lacs, where $8
\lta \delta \lta  23$ and the median value is 16.   In the analysis of
the SED of Centaurus A, we obtained an even lower value, $\delta =1.2$
\citep{cena}.  In that case, we showed that this could be explained by
the presence  of a low  velocity component in  the jet, e.g.   a shear
layer, which  could be the  origin of the observed  nuclear radiation.
This has been interpreted as a further evidence that relativistic jets
have at least  two velocity components: a fast  ``spine'', with plasma
Lorentz  factors  $\Gamma\sim  10-20$   which  dominates  the  BL  Lac
emission, and a slower,  mildly relativistic layer, which accounts for
the emission  of the nuclei  of FR~I radio galaxies.   Considering all
possible  intrinsic  jet velocities,  the  maximum  viewing angle  for
$\delta    =     3.2$    is    $\theta=18^{\circ}$     obtained    for
$\delta=\Gamma=3.2$.  In NGC~6251,  $\delta=3.2$  is still  compatible
with ``typical''  BL Lac Lorentz  $\Gamma$ factors but, in  this case,
the  viewing  angle  to  the  jet  axis would  be  even  smaller  (for
$\Gamma=15$ and $\delta=3.2$ the viewing angle is $\theta = 11^\circ$).

The value for the angle of the line-of-sight to the jet axis we derive
is  rather small, but still  compatible with the  constraint $\theta <
47^{\circ}$ obtained from  VLBI  observations  of  the  jet-counterjet
ratio  \citep{jones02}.   In absence  of a  firm  determination of the
viewing angle to the radio jet axis, we cannot distinguish between the
two scenarios proposed above for the  location of the emitting region,
i.e.   the   fast ``BL   Lac''   component or  a    slower  (but still
relativistic) region of the jet.

Our    estimate  of $\theta$ is  substantially     lower than both the
orientation    of  the   external    dusty   disk ($\theta_{d,ext}\sim
76^{\circ}$) and the inner gas disk ($\theta_{d,inn}\sim 36^{\circ}$),
as modeled    by \citet{ferrarese}.   These authors showed    that the
structure is  warped and tilted ,  and the axis of  the  inner disk is
significantly twisted with      respect  to the  outer    dusty  disk.
Therefore, our  upper limit  $\theta=18^{\circ}$ on the  jet direction
confirms and extends their picture.  The jet axis is not perpendicular
to any  of the observed disks: the  warping of the accretion structure
around the central black holes   would continue towards the  innermost
regions, where the inner accretion  disk and the  jet are likely to be
perpendicular.

The small  viewing   angle $\theta$    we  find  would imply a
deprojected total size of the source $\gta  5.5$ Mpc, which would make
NGC~6251 by  far the largest known  radiogalaxy. However, we point out
that the morphology of its jets, on the largest scales, is complex and
presents  significant bending.     Therefore,   our  determination  of
$\theta$,  which refers to the inner  jet, is probably not appropriate
to estimate the overall deprojected size of this radio source.

\section{Conclusions}

We  have determined the  overall spectral  energy distribution  of the
nucleus  of  NGC~6251,  from the  radio  to  the  gamma ray  band,  by
collecting data from  the literature and analyzing HST  and X-ray data
taken with {\it Beppo}SAX and {\it Chandra}. A significant improvement
with respect to previous work is the homogeneous analysis of HST data,
which span the  range of wavelengths from the IR  ($\sim 2.2 \mu$m) to
the UV ($\sim 2500$  \AA) and allows us to define a  broad peak in the
SED at low energies, and  the analysis of previously unpublished X-ray
data.  The X-ray  data, together  with the  gamma-ray  EGRET detection
recently  published by \citet{egret},  define a  further peak  at high
energies.  The SED is therefore similar  to that of BL Lac objects, in
agreement with the AGN unification model.

Before proceeding to model the SED, a key point to be addressed is the
nature of the optical source.  We have shown that in diagnostic planes
for  the nuclear  emission  of radio  galaxies,  NGC~6251 follows  the
general  behavior of  FR~I.  Therefore,  as  for the  large
majority of 3CR FR~I radiogalaxies, the unresolved nucleus seen in HST
images at  all wavelengths can be interpreted  as synchrotron emission.
At high energies, the {\it  Beppo}SAX hard ($\alpha = 0.78$) power-law
spectrum and  the gamma ray  radiation are most plausibly  produced by
the inverse-Compton process. 

In  light  of the  above  results,  we have  modeled  the  SED in  the
framework of a synchrotron  self-Compton scenario, which well accounts
for  the emission  of blazars  in general,  and especially  of  BL Lac
objects. The parameters  of the model are very  similar to those found
in BL Lacs,  except for the beaming factor,  as qualitatively expected
by  the  unification models.   More  quantitatively,  $\delta \sim  3$
implies that  the viewing angle to  the jet axis cannot  be wider than
$\theta  \sim 20^{\circ}$,  in  agreement with  the radio  information
which claims for a small viewing angle.

In  summary, our  analysis of  the  overall SED  of NGC~6251  directly
supports  the  FR~I--BL Lac  unification  scheme  and strengthens  the
results we  have obtained for the closest  FR~I, Centaurus~A. However,
it  is still  possible   that  the radiation  which  dominates BL  Lacs
emission  is  not produced  co-spatially  with  the nuclear  radiation
observed  in radio-galaxies. Because  of the  larger viewing  angle to
these latter sources, it is plausible that the ``blazar'' component is
beamed  away from  the  observer's  line of  sight,  and the  observed
radiation is produced in a slower  region of the jet, possibly a shear
layer, as proposed for other sources \citep{pap3,cena}.

\begin{acknowledgements}

This work has been partly supported by an ESA post-doctoral
fellowship.  The authors thank the referee for insightful comments.
MC whishes to thank G. Ghisellini for useful discussion.

\end{acknowledgements}

\appendix

\section{SSC model}
\label{appendice}
The model  is described in  detail in \citet{cg} and  \citet{cena}, in
which it has been applied for  the first time to reproduce the nuclear
SED  of a  radio  galaxy,  Centaurus~A.  Here  we  summarize the  main
assumptions of the model.

The source  is a spherical  homogeneous region, embedded in  a tangled
magnetic field.  Relativistic electrons are continuously injected at a
rate  $Q(\gamma)$ [cm$^{-3}$  s$^{-1}$] $\propto  \gamma^{-p}$ between
$\gamma_{\rm min}$ and $\gamma_{\rm  max}$ ($\gamma$ being the Lorentz
factor), and they loose their energy radiatively.
 
The  parameters of  the model  are: the  size of  the source  $R$, the
magnetic  field  $B$,  the  injected  luminosity  $L_{\rm  inj}$,  the
relativistic beaming factor $\delta=[\Gamma(1-\beta \cos\theta)]^{-1}$
(where $\theta$  is the  angle between  the jet axis  and the  line of
sight), $\gamma_{\rm min}$, $\gamma_{\rm  max}$ and the slope $p$. The
resulting electron distribution at  equilibrium is a broken power-law.
Electrons  are  also  allowed  to  escape the  emitting  region  (thus
reducing  the  particle  density,  mimicking source  expansion)  in  a
timescale $t_{\rm esc}$.

We obtain  the equilibrium solution  of the continuity  equation which
governs the  temporal evolution of the  emitting electron distribution
$N(\gamma,t)$.
\begin{displaymath}
\frac{\partial N(\gamma,t)}{\partial t} = \frac{\partial}{\partial\gamma} 
\left[ \dot\gamma(\gamma,t) N(\gamma,t)\right] + Q(\gamma,t) - 
\frac{N(\gamma,t)}{t_{esc}}=0
\end{displaymath}
where  $\dot\gamma= \dot\gamma_{\rm  s} +  \dot\gamma_{\rm C}$  is the
total  (synchrotron  +  self--Compton)  cooling rate.   We  solve  the
equation  numerically  (see  Chiaberge  \&  Ghisellini  1999)  and  we
calculate self-consistently the produced equilibrium spectrum. We take
into  account  the effects  of  the  Klein-Nishina  decline, using  an
approximated step function \citep{zdziarski}.

\clearpage


\clearpage

\clearpage

\end{document}